\documentclass{iaus}
\usepackage{graphicx}

\title[IAU~253.~~The Rise of the Vulcans] 
{The Rise of the Vulcans\footnote{For an examination of an entirely different group of Vulcans
that also made its first appearance in 1999, see the book by James Mann of the same title (Mann 2004).}}

\author[David Charbonneau]   
{David Charbonneau$^1$}

\affiliation{$^1$Harvard-Smithsonian Center for Astrophysics, \\ 60 Garden Street,
MS-16, Cambridge, MA 02138 USA \\ email: {\tt dcharbonneau@cfa.harvard.edu}}

\pubyear{2008}
\volume{253}  
\pagerange{1--8}
\jname{Transiting Planets}
\editors{F.\ Pont et al., eds.}
\begin{document}

\maketitle

\begin{abstract}
In this introductory review presented at the IAU Symposium 253 ``Transiting Planets'', I summarize the
path from the initial 1995 radial-velocity discovery of hot Jupiters to the current rich 
panoply of investigations that are afforded when such objects are observed to transit their parent stars. 
Forty transiting exoplanets are now known, and the time for that population to double
has dropped below one year.  It is only for these objects that we have direct estimates of
their masses and radii, and for which (at the current time) we can undertake direct studies 
of the chemistries and dynamics of their atmospheres. Informed by the successes of hot Jupiter
studies, I outline a path for the spectroscopic study of certain habitable exoplanets 
that obviates the need for direct imaging.
\end{abstract}

\firstsection 
\section{Introduction: The Era of Comparative Exoplanetology}
Although the possibility of detecting planets that transit their parent stars had been
described decades before (Struve 1952; Rosenblatt 1971; Borucki \& Summers 1984), it was not
until the discovery of hot Jupiters by the radial-velocity method (Mayor \& Queloz 1995;
Butler et al.\ 1997) that dedicated transit surveys were begun in earnest.  The justification was
straightforward:  Whereas Jupiter analogs would present a signal only once every 12 years and
have a geometric transit probability of 0.1\%, hot Jupiters orbiting at 0.05~AU
from their stars have orbital periods of roughly 3 days and geometric transit probabilities of 10\%.  
The first success came in 1999 with HD~209458b (Charbonneau et al.\ 2000;
Henry et al.\ 2000; Mazeh et al.\ 2000), but this object was first identified by radial-velocity monitoring and
subsequently determined to transit.  A barrage of ground and space based transit surveys ensued
(see review by Charbonneau et al.\ 2007) yet it was not until 2003$-$2004 that the OGLE survey 
delivered the first handful of exoplanets identified by transits (e.g.\ Udalski et al. 2002; Konacki et al.\ 2003;
Bouchy et al.\ 2004).  Although the relative importance
of the OGLE planets may arguably diminish in the future (since they orbit much more distant stars
and hence follow-up observations are difficult or precluded), they played an absolutely pivotal role
in quenching the thirst of the community after the 2000-2002 drought.
The first wide-field survey (geared toward identifying transiting planets of nearby stars)
to deliver a discovery was TrES, with the 2004 discovery of the aptly named TrES-1 (Alonso et al.\ 2004).  By 2006,
three additional surveys (XO, McCullough et al.\ 2006; HAT, Bakos et al.\ 2007; and SuperWASP,
Collier Cameron et al.\ 2007) produced their first detections and all 3 are now
operating in production-line mode with numerous new planets in the past 18 months.  Fig.\ 1
makes clear the very recent rise in the rate of discovery.  Thus, this conference
convened at a very special moment in history, when the doubling time for new
discoveries of transiting planets (40 as of 31 July 2008) first dipped below one year.  
A little over two years ago I co-authored a review (Charbonneau et al.\ 2007) of the 9 transiting exoplanets known at 
the time, and we concluded that our paper would surely be eclipsed by the coming rapid pace of new discoveries.  
It has been a rare joy to see a publication rendered obsolete so quickly.

My personal excitement over transiting planets stems from three distinct opportunities that
they present:  First, transiting planets permit us to determine their masses and radii
in a manner that is nearly free of astrophysical assumption.  Those estimates, in
turn, bear upon our knowledge of the bulk composition and physical structure of these worlds,
and surely constrain models of their formation (just imagine how our understanding
of the Solar system would differ if he had no knowledge of the bulk composition or
true masses of its planets!).  Second, by observing
the modulation of the combined light of the planet and star as the two undergo mutual
eclipses, astronomers have devised cunning techniques to study the chemistries
and dynamics of their atmospheres \textit{without} the need to image the planet directly.
Third, I feel that it is these methods, which I will call the dynamics-based approach,
rather than high-contrast-ratio imaging, that will permit the first studies of the
compositions of habitable planets and their atmospheres.  A particularly
attractive opportunity exists for such planets orbiting low-mass M-dwarf stars.  In
the sections below, I will consider each of these three motivations in modestly more
detail and provide a sampling of recent results, leaving a thorough and detailed
review to my Vulcan colleagues in their respective chapters of the book that follows.

\begin{figure}[htb]
\begin{center}
 \includegraphics[width=5.4in]{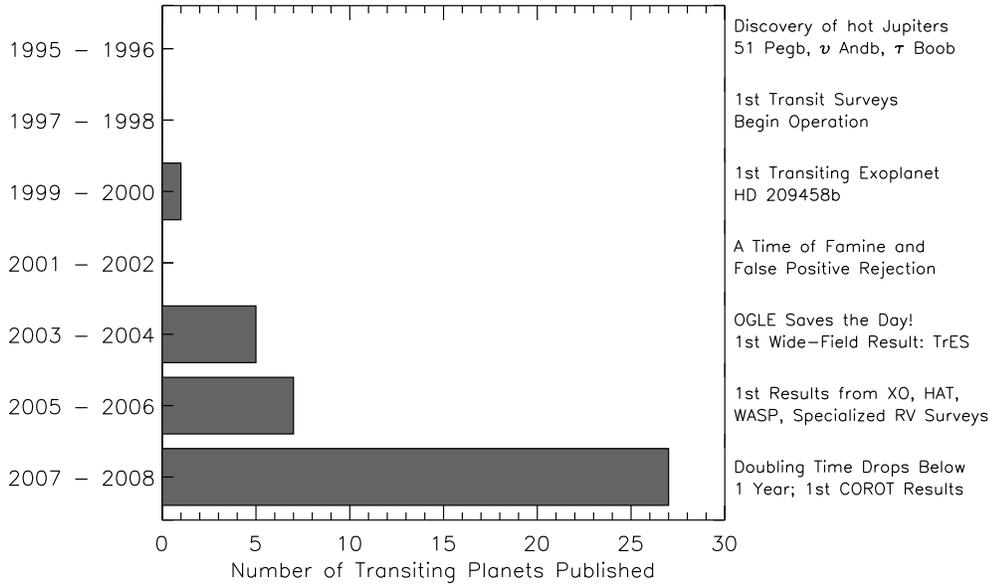} 
 \caption{Histogram of the number of transiting exoplanets (as of 31 July 2008) with confirming radial velocity measurements
as a function of the year that the discovery paper that had been accepted for publication
in a refereed journal was released to the general public.  An incomplete list of notable events is given
on the right hand side.  Ground-based transit surveys were not pursued in earnest until after the 
radial-velocity discovery of hot Jupiter exoplanets in 1995$-$1996, since hot Jupiters present a geometric
probability of a transit that is 100$\times$ greater than that of a planet orbiting at the
Jupiter-Sun separation.  As this plot shows only those planets meeting the publication
criterion as of mid-2008, the 2007$-$2008 bar is incomplete.  At the current epoch, the time for the
number of transiting exoplanets to double is less than 1 year.}
   \label{fig1}
\end{center}
\end{figure}
\newpage

\section{Constraints on Exoplanet Physical Structures and Atmospheres}
Given a mass for the primary star, transiting planets permit the determination of the planetary
masses and radii without additional assumption.  As a result (and after much hard work), we now have an
observational mass-radius diagram for planets (Fig.\ 2).  These estimates
provide the first direct constraints on models (e.g. Burrows et al.\ 2007; Fortney et al.\ 2007)
of the physical structure of gas and ice exoplanets (Fig.\ 3).  While for some of the objects
the agreement between the observed radii and the values predicted for the observed mass and an
assumed composition are in good agreement, there exist many planets with radii that greatly exceed the
model predictions.  Although excellent ideas have been proposed to explain the inflated radii
(see Charbonneau et al.\ 2007 for a summary), none are as of yet satisfying:  Those models
with clear observational consequences (e.g. the ongoing dissipation of orbital eccentricity)
have been ruled out for many systems, and no surviving model yet explains the \textit{diversity}
that is observed, nor dares to predict (based on external measurables, such as the
planet-star separation, stellar luminosity, and stellar metallicity) \textit{which} planets will
be puffy.

\begin{figure}[htb]
\begin{center}
 \includegraphics[width=5.0in]{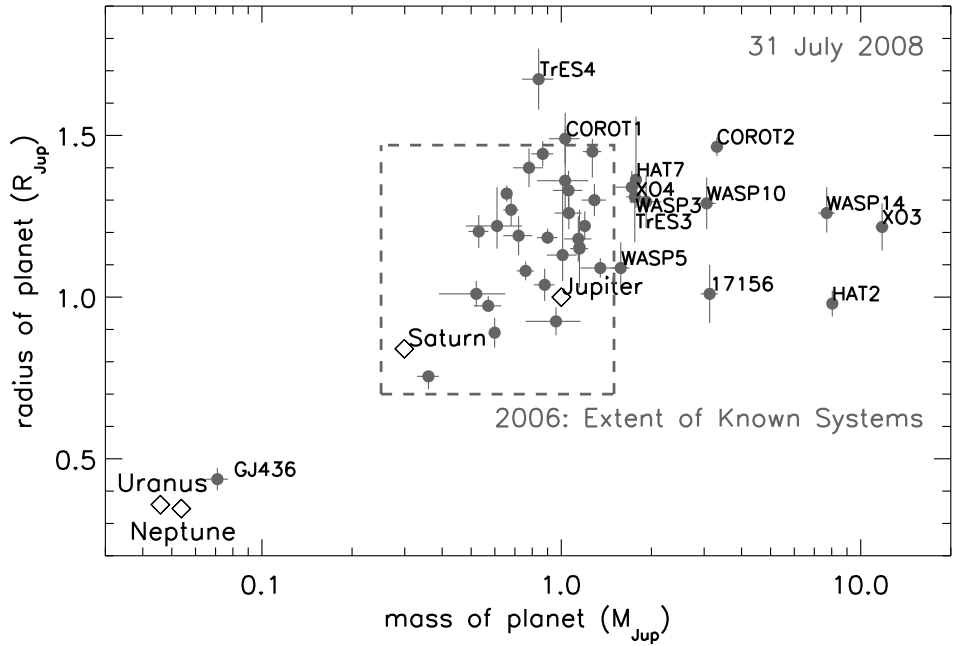} 
 \caption{Masses and radii for the known transiting planets meeting the publication criterion described 
in the caption to Fig.~1.  The Solar system giants are shown for comparison. The
box indicated by the dashed line encloses the 13 transiting planets announced as of December 2006.
Of the 27 transiting planets announced since then, 13 fall inside the box and 14 fall outside it.  The names
of the latter objects are indicated.  Intriguingly, the new discoveries have served to increase greatly the span of physical properties,
indicating that the overall diversity may not yet be mapped.  Two classes of objects merit
special note:  First, there exists a population of transiting companions with masses greater than 1.5$\times$
that of Jupiter and yet below the cut-off of $13~M_{\rm Jup}$ (above which these objects are traditionally
designated brown dwarfs and ultimately low-mass stars).  None of these objects were identified prior to 2007,
despite the fact that the amplitude of the radial-velocity signal is significantly 
greater (and hence confirmation should have been easier) than that for the planets published earlier.  
The second class of note has only one member:
the hot Neptune GJ~436b (Butler et al.\ 2004; Gillon et al.\ 2007), for which the H/He envelope is only
a minority constituent of the planet by mass.}
   \label{fig2}
\end{center}
\end{figure}

\begin{figure}[htb]
\begin{center}
 \includegraphics[width=5.0in]{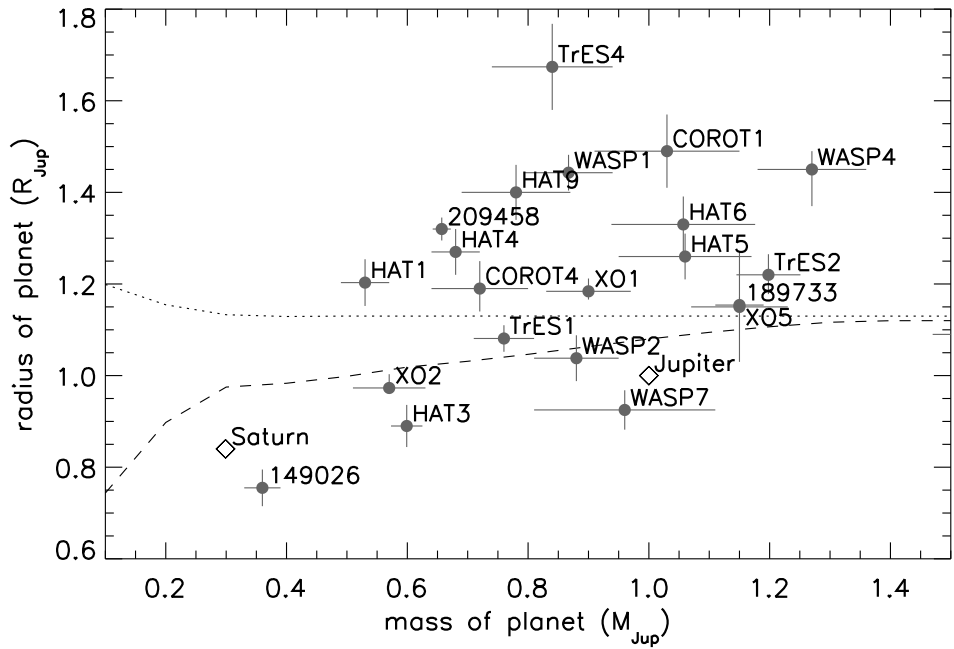} 
 \caption{Masses and radii for the known transiting planets with masses greater than that of Saturn and less than 
1.5$\times$ that of Jupiter.  Jupiter and Saturn are shown for comparison.  I have excluded the OGLE planets as
the estimates of their physical parameters are generally (though not universally) 
less precise than those for the rest of the population (the OGLE planets are included in Fig.~2).
Unlike the mass-radius relation for main-sequence stars, there is no simple observational relationship between these
two quantities for exoplanets.  The dotted line corresponds to the insolated coreless structural models
of Bodenheimer et al.\ (2003) for an age of 4.5~Gyr and a planetary effective temperature of 1500~K.  The dashed line
shows their models for the same parameters but including the presence of a 20~$M_{\oplus}$ core of solid material. 
Insolation alone is insufficient to account for the large radii of many of these planets, and the member of the population
that is at greatest odds with the models is TrES-4, with an average density of $0.22\ {\rm g\, cm^{-3}}$ (Mandushev et al.\ 2007).
HD~149026b (Sato et al.\ 2006), HAT-P-3b (Torres et al.\ 2007), and WASP-7b (Hellier et al.\ 2008) 
all fall significantly below the dashed line, which presumably indicates an enhancement of elements other than H and He.}
\label{fig3}
\end{center}
\end{figure}

Transiting exoplanets have enabled the spectroscopic study of their atmospheres by two distinct methods: First, the technique of transmission
spectroscopy (Charbonneau et al.\ 2002) differences spectra of the star gathered in and out of transit
to reveal wavelength-dependent absorption features indicative of specific atomic and molecular species.  
In the case of HD~189733b, this method recently yielded the detection of water (Tinetti et al.\ 2007),
methane (Swain et al.\ 2008), and the likely presence of small-particle clouds or hazes (Pont et al.\ 2008).
The technique of secondary eclipse observations at infrared wavelengths (Charbonneau et al.\ 2005; Deming et al.\ 2005; sometimes referred
to as occultation spectroscopy) has borne tremendous fruit owing to the remarkable stability of the \textit{Spitzer Space Telescope} (resulting
in part from its heliocentric orbit).  The results are too many to enumerate here, and so
I elect to simply show a representative result in Fig.\ 4.  The study of the dynamics of these strongly-irradiated
atmospheres has become an observational science as a result of \textit{Spitzer}, which permits continuous monitoring of the infrared brightness
of the planet and star for the majority of a planetary orbit.  By inverting the observed changes in brightness, astronomers have
inferred the distribution of temperature with longitude from the substellar point (e.g.\ Knutson et al.\ 2007).

\begin{figure}[htb]
\begin{center}
 \includegraphics[width=2.27in]{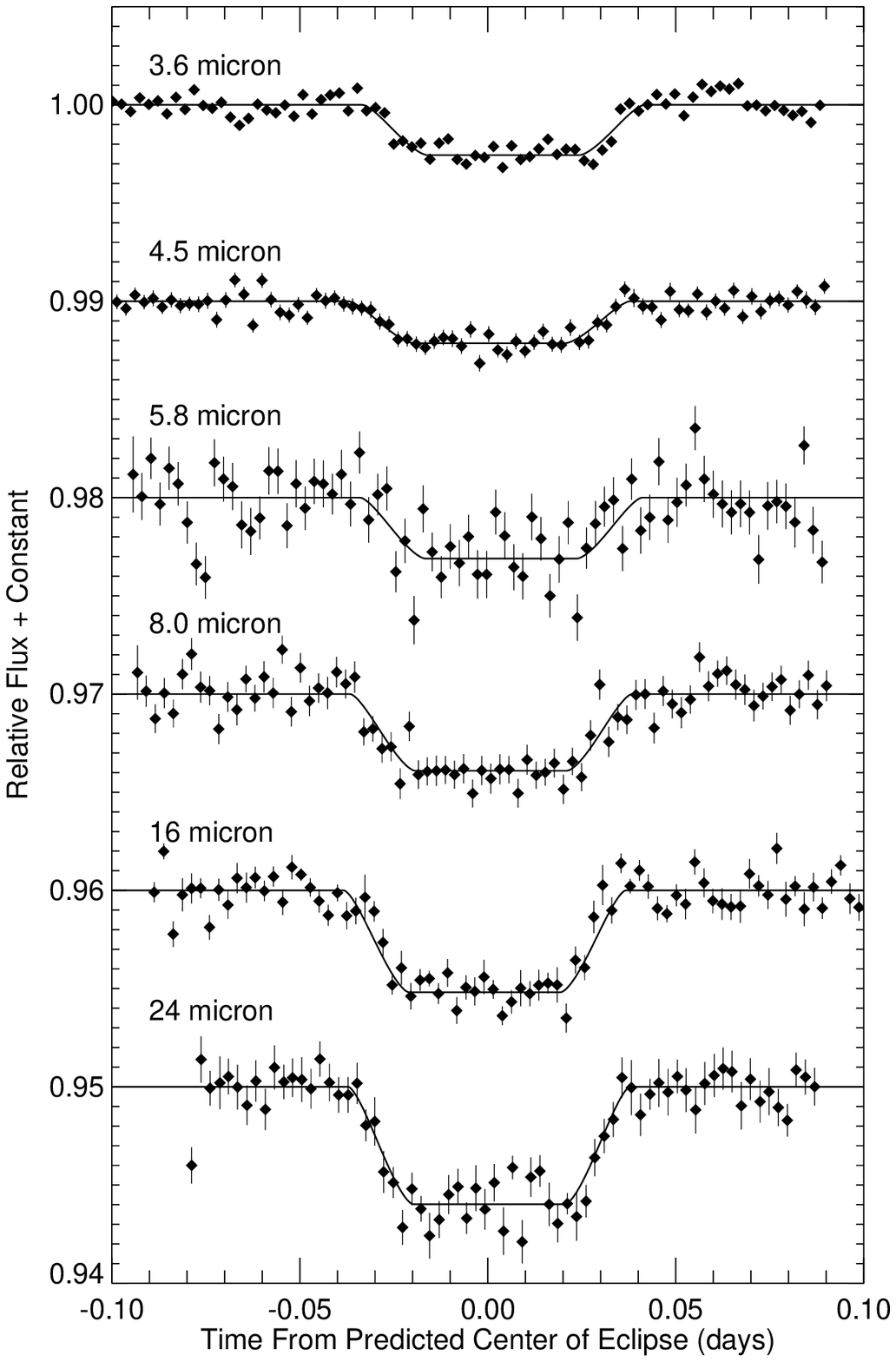}
  \includegraphics[width=3.0in]{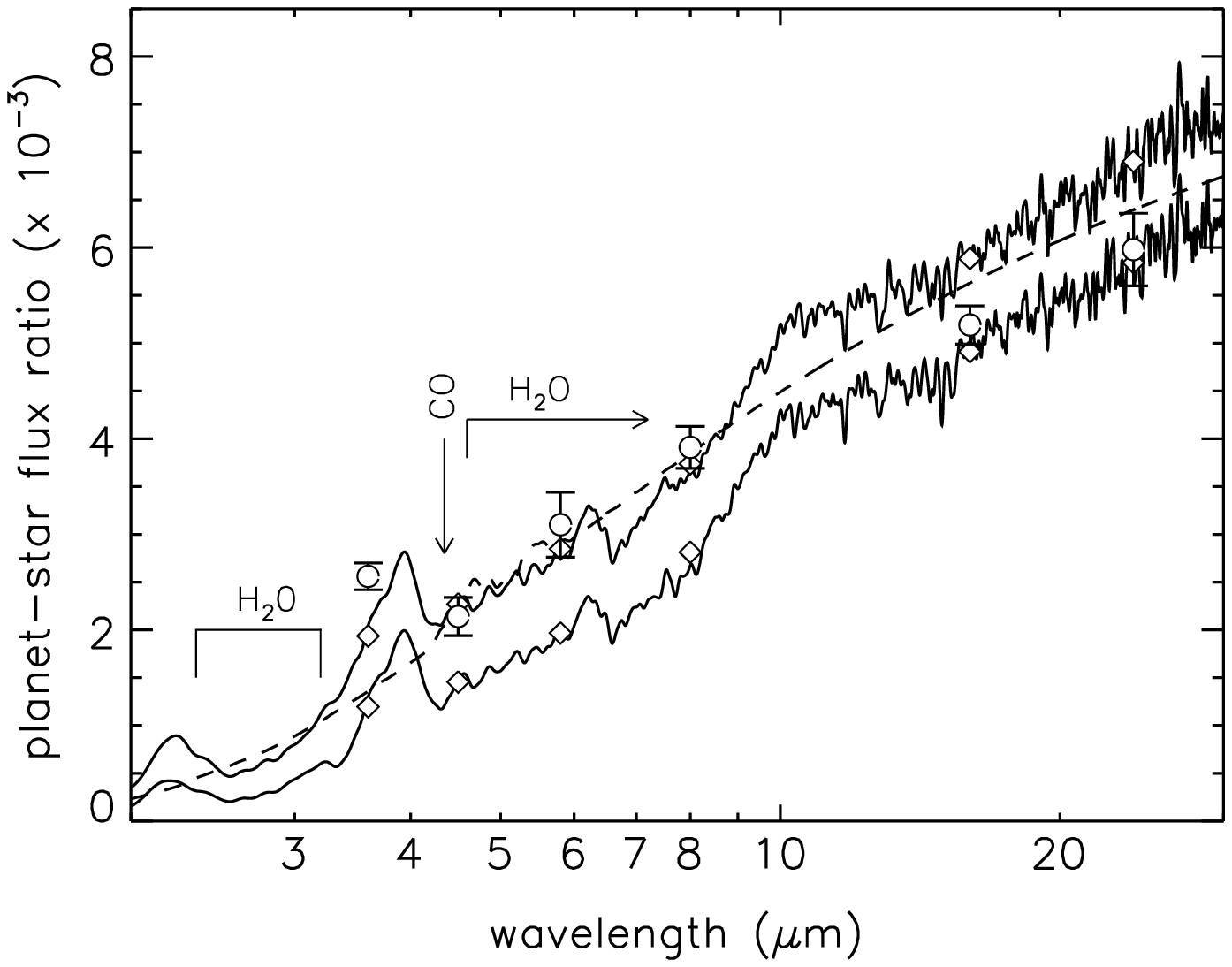}
 \caption{\textit{(from Charbonneau et al.\ 2008)} Left panel: Time series photometry of HD~189733 spanning times of
secondary eclipse when the planet passes out of view behind the parent star.  The time series have been corrected
for detector effects, including intra-pixel sensitive variations and an illumination-dependent ramp in sensitivity.   
All of the data were obtained with the \textit{Spitzer Space Telescope}:  The 3.6, 4.5, 5.8, and 8.0~$\mu$m light curves used the IRAC instrument,
and the 24~$\mu$m light curve was gathered with the MIPS instrument.  The 16~$\mu$m data were gathered by Deming et al.\ (2006)
using the IRS instrument. The best fit eclipse curves are overplotted.  Right panel:  Measurements of the relative depths of the eclipses in each band pass
serve as estimates of the color dependent planet-to-star flux ratios.  These estimates are shown as circles with error bars, and
are compared to models (Barman 2008) of the planet-to-star flux ratio under the assumption that the emission of the absorbed stellar flux 
is constrained to only the day side (upper curve), or redistributed uniformly over the entire planet (lower curve).  The predicted
ratios in the \textit{Spitzer} band passes (obtained by integrating these models over the band pass response curves) are plotted as
diamonds.  The flux ratio under the assumption that the planet radiates a Planck spectrum with a temperature of
1292~K (the best-fit value) is shown as a dashed line.  The dashed line is a poor fit, indicating that we have detected
spectral variations (primarily due to water and CO) that are broadly in agreement with the model predictions.}
\label{fig4}
\end{center}
\end{figure}

\newpage

\section{The Small Star Opportunity}
In the next few years, one of the most exciting opportunities will be to export the techniques developed for
the observation of gas giants transiting F, G, and K dwarf stars to the study of terrestrial planets transiting
M dwarf stars (e.g.\ Charbonneau \& Deming 2007; Gaidos et al.\ 2007; Nutzman \& Charbonneau 2008).  
The advantages are particularly compelling for the study of such planets orbiting within their
stellar habitable zones: Consider the case of a 2$-R_{\oplus}$ radius planet orbiting at 1~AU from a G2V star,
and compare it to that of a planet of the same size orbiting an M5V star (0.25~$M_{\odot}$, 0.25~$R_{\odot}$, 0.0055~$L_{\odot}$) 
at the distance such that it would receive the same energy per unit surface area and unit time.  The sizes of
the two orbits are drawn to scale in Fig.\ 5.  

\begin{figure}[htb]
\begin{center}
 \includegraphics[width=4.in, angle=270]{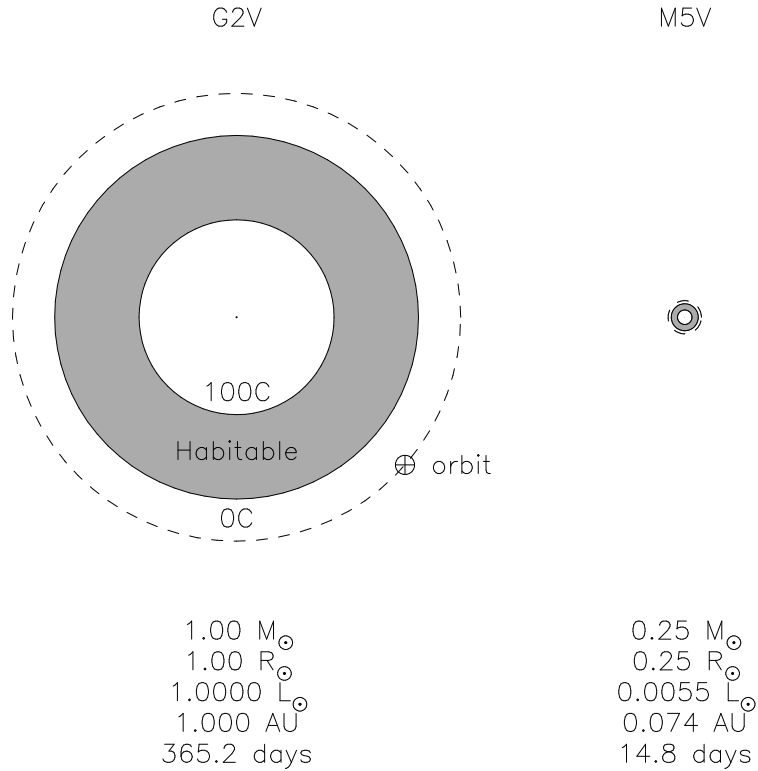} 
 \caption{\textit{(Credit: J.\ Irwin)} The shaded regions denote the range of distances 
from a G2V star (left) and an M5V star (right) for which the equilibrium
temperature of the planet is greater than 0$^{\rm o}$C and less than 100$^{\rm o}$C,
and hence water might be liquid at the surface.  This naive definition
ignores the greenhouse effect, which maintains the surface temperature of the
Earth roughly $+30^{\rm o}$C above the equilibrium temperature.  The Earth's
orbit is indicated by the dashed circle and the orbit at which a planet would
receive the same amount of energy per unit area and unit time is shown as a dashed
circle in the right plot.  For direct imaging studies, the tight habitable zones
of M5V stars present an enormous challenge due to the reduction in the angular separation
of the planet and star.  In contrast, the M5V system is very favorable
for transit surveys, since the transits are more probable and more frequent, and
both the photometric and radial-velocity signals are much larger than they would
be for the G2V system.  The planet-to-star contrast, which depends upon the relative
surface areas and brightness temperatures of the planet and star, is much larger for the M5V 
primary compared to the G2V system, facilitating the measurement of the planetary spectrum
by occultation spectroscopy (see Fig.\ 6).}
\label{fig5}
\end{center}
\end{figure}

For a transit hunter, the M5V planet presents numerous advantages.  First, transits would be
\textit{more likely}:  The M5V planet would present a 1.6\% geometric probability of a transit, 3.2$\times$ greater than the 
value for the G2V planet.  Second, transits would be \textit{more frequent}:  At only 0.074~AU from the M5V, the planet would
orbit once every 15 days as opposed to 1 year.  Third, transits would be a \textit{much larger signal}:  The small
radius of the M5V dwarf means that the planet would present a transit depth of 0.5\% as opposed to 0.03\% for a G2V primary.
Fourth, the small mass of the M5V and the short orbital period serve to \textit{boost the amplitude of the
stellar radial-velocity signal}, facilitating a confirmation of the planet and the determination of its mass:  The peak-to-peak
amplitude of a 7~${M_\oplus}$ planet would be $10\ {\rm m\, s^{-1}}$ for the M5V star, whereas this would shrink to
$1.3\ {\rm m\, s^{-1}}$ for the Sun-like primary.  These considerations have driven a number of radial-velocity surveys
to lavish attention on M-dwarfs, and are the basis for the MEarth Project (Nutzman \& Charbonneau, Irwin et al.\ 2008),
a dedicated photometric transit survey that will use 8 40-cm telescopes to survey 2000 northern stars with radii
smaller than 0.33~$R_{\odot}$ and at distances less than 33~pc from the Sun.

The habitability of M-dwarf planets has recently been re-examined by Tarter et al.\ (2007) and Scalo et al.\ (2007).  Work
reviewed therein contests the conclusions of earlier studies, which found that such planets, which are likely to
be tidally-locked to their stars, would be inhospitable to life due either to atmospheric collapse or steep day-night
temperature gradients.  Preliminary studies (see Fig.\ 6) indicate that the atmospheres of such planets would
be accessible to spectroscopic study using the \textit{James Webb Space Telescope}.  Rather than separating the light of the planet from that
of the star \textit{spatially} through high-contrast ratio imaging, we would separate them \textit{temporally} using the technique of
occultation spectroscopy.  For the planet considered earlier,
the planet-to-star flux ratio is 0.05\% (in the Rayleigh-Jeans limit) for the M5V primary but only 0.0017\% for the G2V star.
The former level is above that achieved with the \textit{Spitzer Space Telescope}.  It bears noting that
\textit{Spitzer} itself could undertake studies of the thermal emission from terrestrial planets orbiting M-dwarfs,
but only if such planets are significantly closer (hence hotter) than the nominal habitable zone.  The preeminence of \textit{JWST}
for such work relies on both its stability and its sensitivity at wavelengths greater than $10~\mu$m, extending into the Rayleigh-Jeans
limit for a planet at 300~K.  Provided such habitable-zone M-dwarf planets exist, they present the least arduous route to undertaking 
the spectroscopic search for biomarkers in the atmosphere of a planet orbiting another star.
\begin{figure}[htb]
\begin{center}
\includegraphics[width=4.5in]{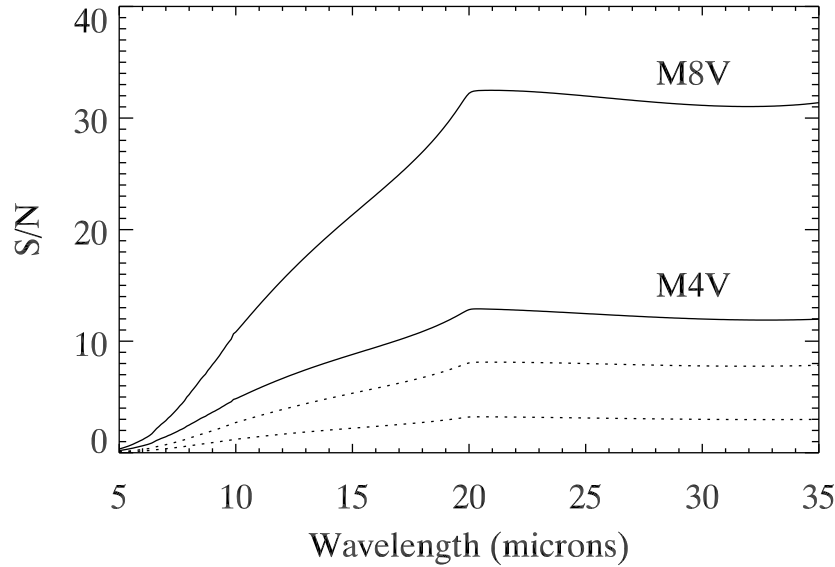} 
 \caption{\textit{(Credit: D.\ Deming)} The signal-to-noise ratio that could be obtained
on the emission spectrum (assuming a spectral resolution of 40) of an exoplanet using the technique of occultation spectroscopy
and the MIRI instrument (Wright et al.\ 2004; Wells et al.\ 2006) aboard the \textit{James Webb Space Telescope} with 200 hours of integration.  The upper solid
line assumes a planet with a temperature of 300~K and a radius of twice that of the Earth orbiting an M8V primary.
The lower solid line shows the prediction for the same planet orbiting an M4V primary.  The dotted lines
show the results for the same pair of primaries but reducing the planet radius to the terrestrial value.  These results
indicate that with a substantial investment of time, we could use \textit{JWST} to detect broad band absorption features
due to molecules in the atmosphere of a habitable planet, provided we are successful in identifying
terrestrial planets orbiting within the habitable zones of late M dwarfs.}
\label{fig6}
\end{center}
\end{figure}

\clearpage
\acknowledgments I would like to thank current and recently-graduated students S.\ Ballard, C.\ H.\ Blake, J.\ Devor,
H.\ A.\ Knutson, F.\ T.\ O'Donovan, and P.\ Nutzman, and post-doctoral fellows J.\ L.\ Christiansen, J.\ Irwin, and D.\ T.\ F.\ Weldrake
for the stimulating daily discussions that have made the Harvard-Smithsonian Center for Astrophysics such an enjoyable 
and productive place for exoplanet research.

\end{document}